\documentclass[reprint,
longbibliography,
amsmath,amssymb,
 aps]{revtex4-1}

\usepackage{graphicx}% Include figure files
\usepackage{dcolumn}% Align table columns on decimal point
\usepackage{bm}% bold math

\usepackage{natbib}

\usepackage{verbatim} 
\usepackage{multirow} 
\usepackage{graphics}
\usepackage{subfigure}
\usepackage{array}
\usepackage{ulem}

\usepackage{epsfig}

\begin{document}

\preprint{}

\title{Comparative numerical studies of ion traps with integrated optical cavities}

\author{Nina Podoliak$^1$, Hiroki Takahashi$^2$, Matthias Keller$^2$, Peter Horak$^1$}

\affiliation{$^1$ Optoelectronics Research Centre, University of Southampton, Southampton SO17 1BJ, U.K.\\
$^2$ Department of Physics and Astronomy, University of Sussex, Falmer, BN1 9QH, U.K.}

\date{\today}% It is always \today, today,

\begin{abstract}
We study a range of radio-frequency ion trap geometries and investigate the effect of integrating dielectric cavity mirrors on their trapping potential using numerical modelling. We compare five different ion trap geometries with the aim to identify ion trap and cavity configurations that are best suited for achieving small cavity volumes and thus large ion-photon coupling as required for scalable quantum information networks. In particular, we investigate the trapping potential distortions caused by the dielectric material of the cavity mirrors in all 3 dimensions for different mirror orientations with respect to the trapping electrodes. We also analyze the effect of the mirror material properties such as dielectric constants and surface conductivity, and study the effect of surface charges on the mirrors. As well as perfectly symmetric systems, we also consider traps with optical cavities that are not centrally aligned where we find a spatial displacement of the trap center and asymmetry of the resulting trap only at certain cavity orientations. The best trap-cavity configurations with the smallest trapping potential distortions are those where the cavities are aligned along the major symmetry axis of the electrode geometries. These cavity configurations also appear to be the most stable with respect to any mirror misalignment. Although we consider particular trap sizes in our study, the presented results can be easily generalized and scaled to different trap dimensions.
\begin{description}
\item[PACS numbers]
37.10.Ty, 33.50.Dq, 82.80.Ms, 34.70.+e.
\end{description}
\end{abstract}

      
\maketitle

\maketitle

%%%%%%%%%%%%%%%%%%%%%%%%%%%%%%%%%%%%%%%%%%%%%%%%%%%%%%%%%%%%%%%%%%%%%%%%%%%%%%%%
\section{Introduction}
\label{sec:intro}

Atomic ions held in radio frequency traps are currently among the most successful implementations of quantum information processing (QIP). 
Their near-ideal isolation from the environment enables long coherence times for the electronic and motional quantum states of trapped ions. Using resonant electromagnetic radiation, those quantum states can be manipulated and interrogated with high fidelity. As a consequence, a set of universal quantum gates and quantum algorithms have been demonstrated successfully (e.g. \cite{Monroe1995,Schmidt-Kaler2003,Gulde2003,Leibfried2003a,Benhelm2008a,Harty2014}) and strings of up to 14 ions have been entangled \cite{Monz2011}. 
All of these achievements, however, employ the ions' joint motion in a single trapping potential to implement collective quantum logic operations.
As the number of ions in a trap increases, so do the challenges in QIP. This is because addressing individual motional modes and individual ions become increasingly more difficult in a longer ion string. Hence ion trap QIP in single axial potentials is not deemed scalable to large numbers of qubits.
To circumvent this problem, Kielpinski \textit{et al.} \cite{Kielpinski2002} proposed to construct a large scale ion-trap quantum computer from interlinked segments of small ion traps. In this architecture ions are shuttled between segments for communication of quantum information while each segment operates on only a small number of ions. Even though there has been notable progress based on this proposal \cite{Seidelin2006,Home2009},
there are still technological challenges to overcome, such as anomalous motional heating \cite{Hite2013}, fast ion shuttling with negligible decoherence \cite{Bowler2012,Walther2012} and the demonstration of multi-ion manipulation in segmented traps.

Another approach to scale up ion-trap based QIP systems is to utilize optical links. 
The efficient generation of single photons with well-controlled shape, frequency, polarization and at a high repetition rate was demonstrated by spontaneous Raman scattering in single trapped ions \cite{Maunz2007, Kurz2013}. This method was used for establishing atom-photon entanglement \cite{Huwer2013} and entanglement between two atoms at a distance \cite{Monroe2007}.
An optical resonator coupled with trapped ions provides an ion-photon interface with high coupling probability for quantum networking \cite{Pellizzari1995,Cirac1997,Kimble2008}. After the early demonstrations of ions localized in optical cavities \cite{Guthohrlein2001,Mundt2002} several important experimental landmarks have been demonstrated, such as the cavity assisted generation of single photons \cite{Keller2004,Barros2009}, the generation of entanglement between a single ion and a single photon \cite{Stute2012} and the heralded entanglement of two intra-cavity ions \cite{Casabone2013}. Despite these successful demonstrations currently cavity-based ion-photon interfaces are limited by the weak interaction between the ions and cavity mode. To achieve high fidelity and highly efficient ion-photon interfaces, the coherent interaction strength between the ion and the cavity must be larger than the decoherence rates of the ion and cavity states. This strong coupling regime can be achieved by reducing the mode volume of the cavity. However reducing the physical volume of an optical cavity around trapped ions is complicated by the fact that the dielectric surfaces of the cavity mirrors can adversely affect the trapping potential \cite{Harlander}.  
While strong coupling has been demonstrated in systems with neutral atoms and collectively with many ions \cite{Herskind2009}, it still remains elusive for a single ion despite many proposed designs and implementations \cite{Brandstatter2013,Cetina2013,Leibrandt2009,Sterk2012}. 
In this regard the development of fiber-based Fabry-Perot cavities \cite{Hunger2010} has offered a new promising perspective for integrating small optical cavities in ion traps. Their reduced size and possibility of tight integration and electrical shielding has the potential to achieve a small cavity mode volume without seriously compromising the trapping stability \cite{Steiner2013,Takahashi2013, Pfister2016}.

Taking the integration of fiber cavities in ion traps as a basic design for the ion-photon quantum interface, in this article, we present detailed systematic numerical studies of the effects of dielectric materials in the vicinity of ion traps. Evolving from the original parabolic Paul traps, currently a range of different trap designs is available. We consider five geometries of rf ion traps with all different orientations of the optical cavities with respect to the trapping electrodes and compare them in terms of stability against the influence of dielectric mirrors on the trapping potential.  We aim to identify the most robust ion trap and optical cavity configurations that would enable both a strong trapping potential for long ion trapping lifetime and a small cavity volume for large ion-photon coupling as required for successful realization of scalable quantum information networks.

The paper contains the following. Geometries of ion traps and cavities are described in Sec.~\ref{sec:geom}. Simulation methods and the characteristics of the traps without cavity are presented in Sec.~\ref{sec:trap}. Sec~\ref{sec:dist-potent-cavity} compares the distortion of the trapping potential in all 3 dimensions caused by the dielectric cavity mirrors in different ion traps and at different cavity orientations. Traps with optical cavities that are not centrally aligned to the trap electrodes are investigated in Se.~\ref{sec:mirr}. The effect of surface charges on the mirror facets and the effect of mirror material properties such as dielectric constants and surface conductivity are considered in Sec.~\ref{sec:charges} and \ref{sec:material}. The general conclusion is given in Sec.~\ref{sec:conclusion}.

\section{Ion trap and cavity geometries}
\label{sec:geom}

\begin{figure*}[ht]
  \centering
  \includegraphics[width=14cm]{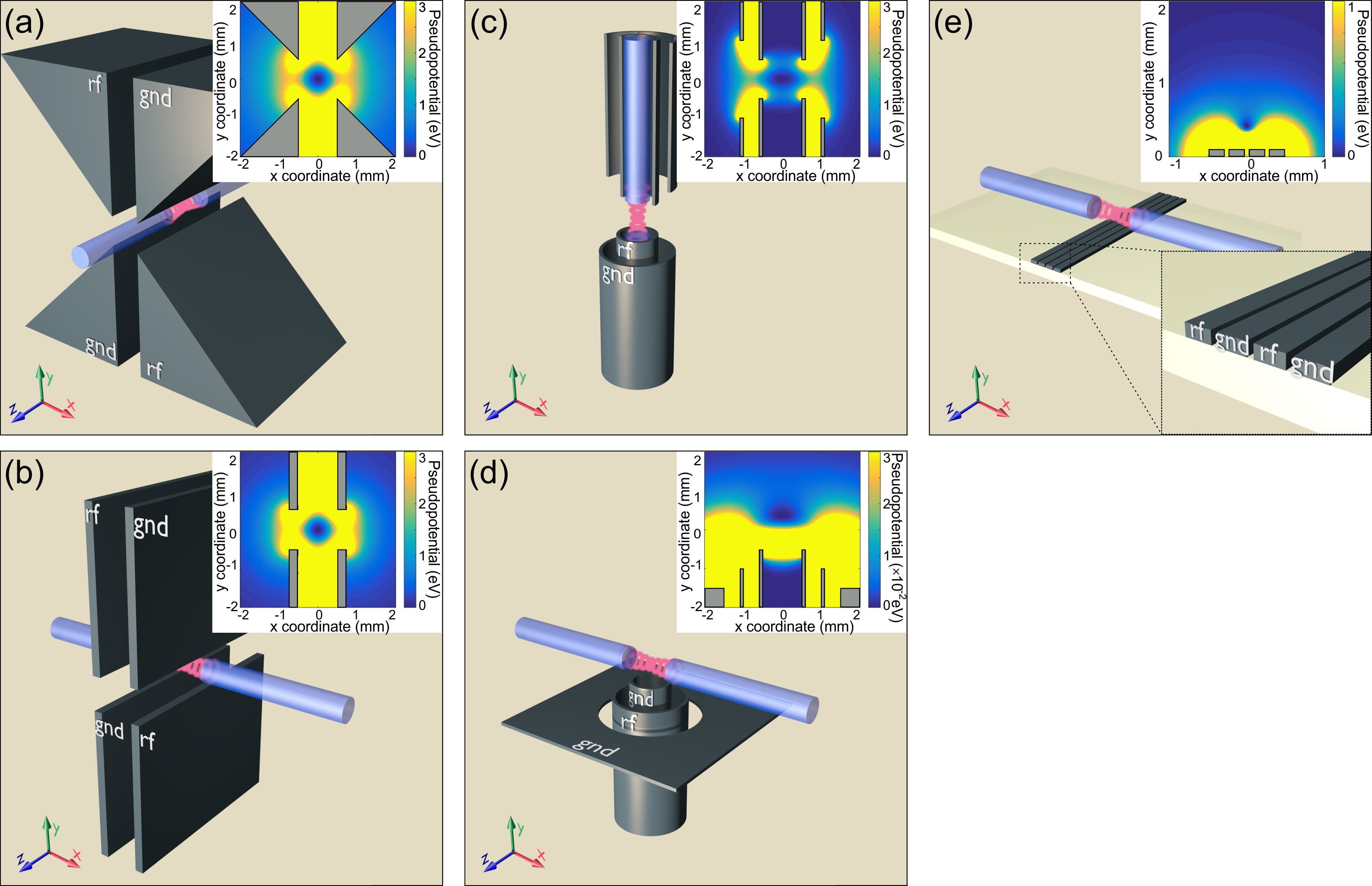}
  \caption{Geometries of ion traps with integrated optical cavities: (a) linear trap with blade shaped electrodes; (b) linear trap with wafer electrodes; (c) endcap trap; (d) stylus trap; (e) surface trap. 'rf' and 'gnd' mark rf- and grounded electrodes, respectively. For each trap geometry different cavity orientations are considered in this article, e.g. with cavities along the $x$-, $y$-, or $z$-axis. The insets show the unperturbed trapping potentials for the respective traps.}
  \label{fig:IonTrap}
\end{figure*}

We consider five different ion trap geometries (see Fig.~\ref{fig:IonTrap}): two linear traps, two cylindrical traps and a planar surface trap. All traps have a typical electrode separation of $1$ mm to allow easy comparison. 
The first linear trap is shown in Fig.~\ref{fig:IonTrap}(a); it consists of four blade-shaped electrodes with an edge angle of $45^\circ$ \cite{Keller2004}.  The second trap, shown in 
Fig.~\ref{fig:IonTrap}(b), has four planar electrodes of $0.25$ mm thickness \cite{Brama2012}. In both traps the horizontal and vertical separations between the electrodes are $1$ mm and the electrode length along the trap axis is $6$ mm. The trapping potential is generated by applying an rf voltage to the two diagonally opposing electrodes while the remaining electrodes are grounded (see Fig.~\ref{fig:IonTrap}).
Usually a set of dc electrodes along the z-axis is used to provide an axial confinement in linear traps. However in this article, we omit these dc electrodes for the sake of simplicity and focus on the effects of the cavity mirrors to the rf potentials in the linear traps. This is because disturbances to a dc potential can be accommodated with compensation dc voltages whereas disturbances to rf potentials are harder to handle and  more critical for the trapping stability.
    
The cylindrical traps are the endcap \cite{Takahashi2013} and the stylus trap \cite{Maiwald2009}. The endcap trap geometry consists of a pair of two concentric hollow cylinders as shown in Fig.~\ref{fig:IonTrap}(c). The inner diameters of the inner and  outer electrodes are $1$ mm and $2$ mm respectively with their thickness being $0.1$ mm. By applying an rf voltage to the inner electrodes while grounding the outer ones, a ponderomotive potential is created with a potential minimum at the center of the electrode gap. The inner electrode protrudes by 0.5 mm from the outer electrode. The vertical inter-electrode distances are $1$ mm and $2$ mm for the inner and outer electrodes respectively. The stylus trap is formed by a single set of concentric cylinders of the same dimensions as the endcap trap with an additional grounded plate (see Fig.~\ref{fig:IonTrap}(d)).  The cylindrical electrodes protrude by 0.5 mm from the ground plane. Applying an rf voltage to the outer cylinder while keeping all the other electrodes grounded creates a ponderomotive potential with a local minimum above the center electrode.

The planar surface trap consists of four parallel electrodes of an identical shape on a flat surface as shown in Fig.~\ref{fig:IonTrap}(e). The electrode height is $0.1$ mm and the width is $0.2125$ mm with gaps between them of $0.05$ mm, which makes the total width of the trap to be $1$ mm. The trapping potential is generated by applying an rf voltage to two interleaved electrodes while grounding the remaining ones. This forms a trapping potential with a minimum above the electrode plane on the center line of the electrode configuration. 

\begin{figure*}[ht]
  \centering
  \includegraphics[width=13cm]{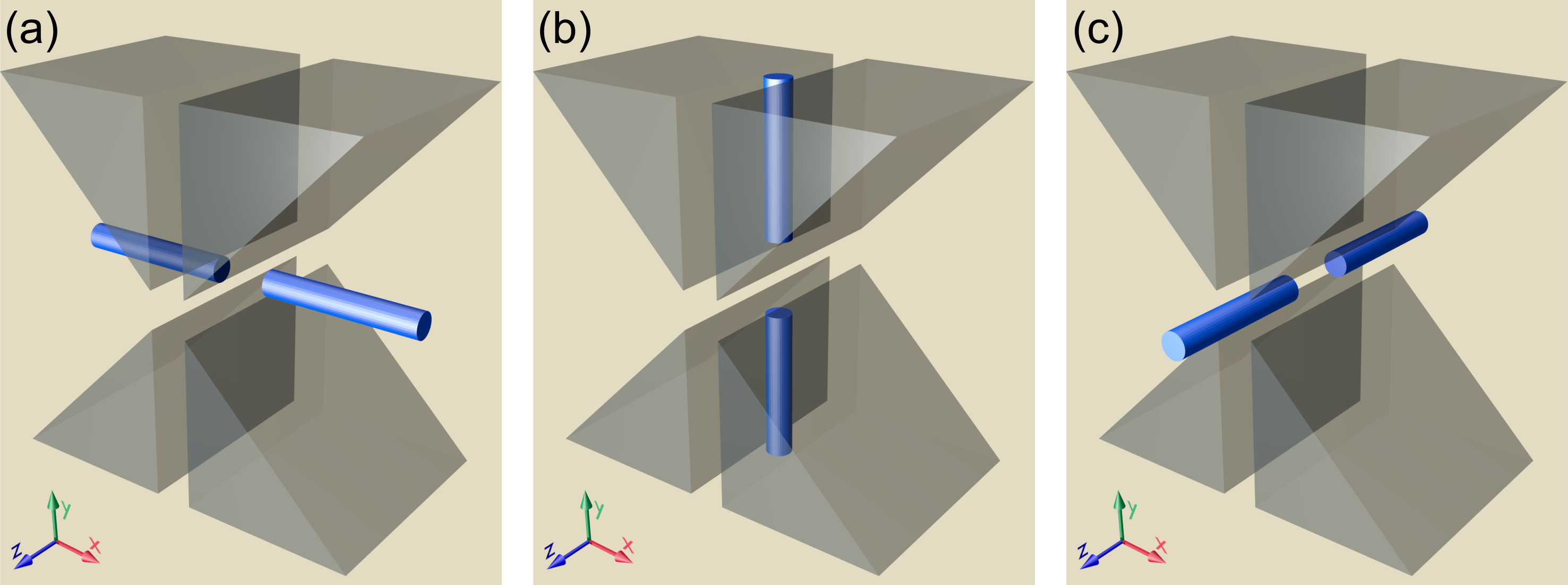}
  \caption{Three orientations of optical cavities integrated in the blade linear trap. Cavities oriented along the (a) $x$-, (b) $y$-, and (c) $z$-axis.}
  \label{fig:Mirr}
\end{figure*}

In all the ion traps, the optical cavity is modeled as a pair of glass cylinders with a diameter of $0.7$ mm and a dielectric constant of $\epsilon=3.8$ for silica glass. The effect of the mirror facets with potentially different materials from the mirror substrate is considered in Sec.~\ref{sec:material}. We investigate different cavity orientations with respect to the trap electrodes. For the linear traps, mirrors can be aligned either along the $x$-, $y$-, or $z$-axis, as shown in Fig.~\ref{fig:Mirr} using the blade trap as an example. For the cylindrical traps, the geometries with mirrors oriented along the $x$- and $z$-axis are equivalent, therefore for these traps only two cavity orientations along the $x$-axis (mirrors from the side to the trap) and along the $y$-axis (mirrors inside the central electrodes) are modeled. For the surface trap, mirrors are aligned either along the $x$-, $y$- or $z$-axis and positioned symmetrically around the rf-nodal line. In the case of mirrors aligned perpendicular to the trap surface (along the $y$-axis), the modeled domain includes only one mirror above the trap surface, as the second mirror would be positioned below the substrate, where the rf-field is screened by the substrate. In this case, the substrate would also have a hole between the two electrodes in the center to allow the formation of an optical mode between the mirrors.

\section{Simulating the trapping potentials}
\label{sec:trap}

Rf-electric field profiles and trapping pseudopotentials created around the trap electrodes are numerically calculated by a finite element method using Comsol Multiphysics$^{\mbox{\textregistered}}$ (AC/DC  module).
In all the cases we assume an rf-voltage amplitude of $200$ V at $10$ MHz. For each trap we determine the calculation domain dimensions at which the outer boundaries do not affect the rf-field in the middle of the trap. Then, the pseudopotential $\Phi_{\mbox{\tiny pseudo}}$ experienced by a trapped ion is calculated from the numerically determined electric field $\vec{E}$ through
\begin{equation}
\Phi_{\mbox{\tiny pseudo}}=\frac{Q}{4M\Omega^2}|{\vec{E}}|^2 ,
\end{equation}
with the ion charge $Q$, its mass $M$ and the angular frequency $\Omega$ of the rf voltage. Throughout this article we use calcium-40 ions with $M = 40 \,au$ and $Q = 1\,e$ where $au$ and $e$ are the atomic mass unit ($= 1.66\times 10^{-27}$ kg) and elementary charge ($= 1.60 \times 10^{-19}$ C) respectively.
Secular frequencies of the ion are obtained by fitting the corresponding pseudopotentials around the trap center by a parabola. A trap depth is extracted as the minimal potential barrier height from the trap center along a particular direction. Table \ref{tab:hresult} contains trap depths and secular frequencies for calcium ions in different trap geometries for the trap parameters given above in the absence of optical cavities. 

\begin{table}
\caption{Trap depths and secular frequencies of ion traps in the absence of optical cavities.} 
\label{tab:hresult}
\begin{ruledtabular}
\begin{tabular}{l r  r r r r r r} 
&\multicolumn{3}{c}{Trap depth (eV)} &\multicolumn{4}{c}{Secular frequency (MHz)}\\  [-1ex]
\raisebox{1.5ex}{Ion trap} &  & x-axis & y-axis & & & x-axis & y-axis\\ [0.5ex] 
\colrule
Blade trap &  & 2.62 & 6.06 & & & 1.23 & 1.23 \\ 
Wafer trap &  & 3.54 & 6.06 & & & 1.32 & 1.32 \\
Endcap trap &  & 1.77 & 0.66 & & & 0.53 & 1.06 \\
Stylus trap &  & 0.03 & 0.005 & & & 0.05 & 0.10\\ 
Surface trap &  & 2.08 & 0.23 & & & 2.43 & 2.44 \\[0.5ex] 
\end{tabular}
\end{ruledtabular}
\end{table}

The linear traps create the deepest potential traps. The traps with the blade and wafer shaped electrodes are characterized by the same trap depths along the $y$-axis, but the trap depth along the $x$-axis for the blade trap is $\sim 25~\%$ smaller than for the wafer trap. The stylus trap creates the weakest ion confinement. The potential barrier of the surface trap is comparable to those of the linear traps along the $x$-axis, but is rather shallow along the $y$-axis. The secular frequencies determined along the $x$- and $y$-axis are the same in the linear and the surface traps, indicating symmetrical shapes of pseudopotentials around the trap center. In the endcap and stylus trap, however, the secular frequencies along the $y$-axis are twice larger than the frequencies along the $x$-axis.

Rescaling the trap geometry changes the trap depth and secular frequencies. When all the dimensions, including the cavities,  are rescaled linearly by a factor of $a$ (for example the electrodes separation of $1$ mm changed to $a$ mm),  the trap depth and secular frequencies change by a factor of $1/a^2$. According to this scaling law, even though we have chosen particular dimensions for the traps, all the results in this article can be rescaled approriately depending on practical design needs.

\section{Distortion of the potentials by the cavity mirrors}
\label{sec:dist-potent-cavity}

The linear and endcap traps create trapping fields with rf minima located at the symmetry axis or point defined by the electrode configurations (see insets in Fig.~\ref{fig:IonTrap}(a)-(c)). An ion is trapped on the central line ($z$-axis) in the linear traps and at the central point between the electrodes in the endcap trap. On the other hand in the stylus and surface traps, a trapped ion is located above the rf-electrodes (see insets in Fig.~\ref{fig:IonTrap}(d) and (e)). For the chosen geometry the stylus trap creates an rf-null point positioned at $0.87$ mm above the edge of the inner electrode and the surface trap creates an rf-null line at a distance of $0.30$ mm above the electrode surface. An advantage of these geometries is a wide access angle for laser addressing to the trapped ions but they create relatively shallow trapping potentials. In the linear and endcap traps, the position of the rf-null is unaffected by the presence of the mirrors when the cavities are aligned symmetrically around it. However, in the stylus and surface traps the rf-null is slightly shifted in the vertical direction even by symmetrically placed dielectric mirrors. For the stylus trap the rf-null is shifted by $30$ $\mu$m and $120$ $\mu$m upwards (away from the electrodes) when the mirrors are aligned along the $x$- and $y$-axis directions respectively at a distance of $0.5$ mm from the unperturbed rf-null. In the case of the surface trap, the rf-null is shifted downwards by $15$ $\mu$m when the mirrors are aligned along the $x$ direction, and upwards (towards the mirror) by either $2.6$ $\mu$m or $2.9$ $\mu$m when the mirrors are aligned along the $y$ or $z$ direction, respectively.

\begin{figure*}[ht]
  \centering
  \includegraphics[width=13cm]{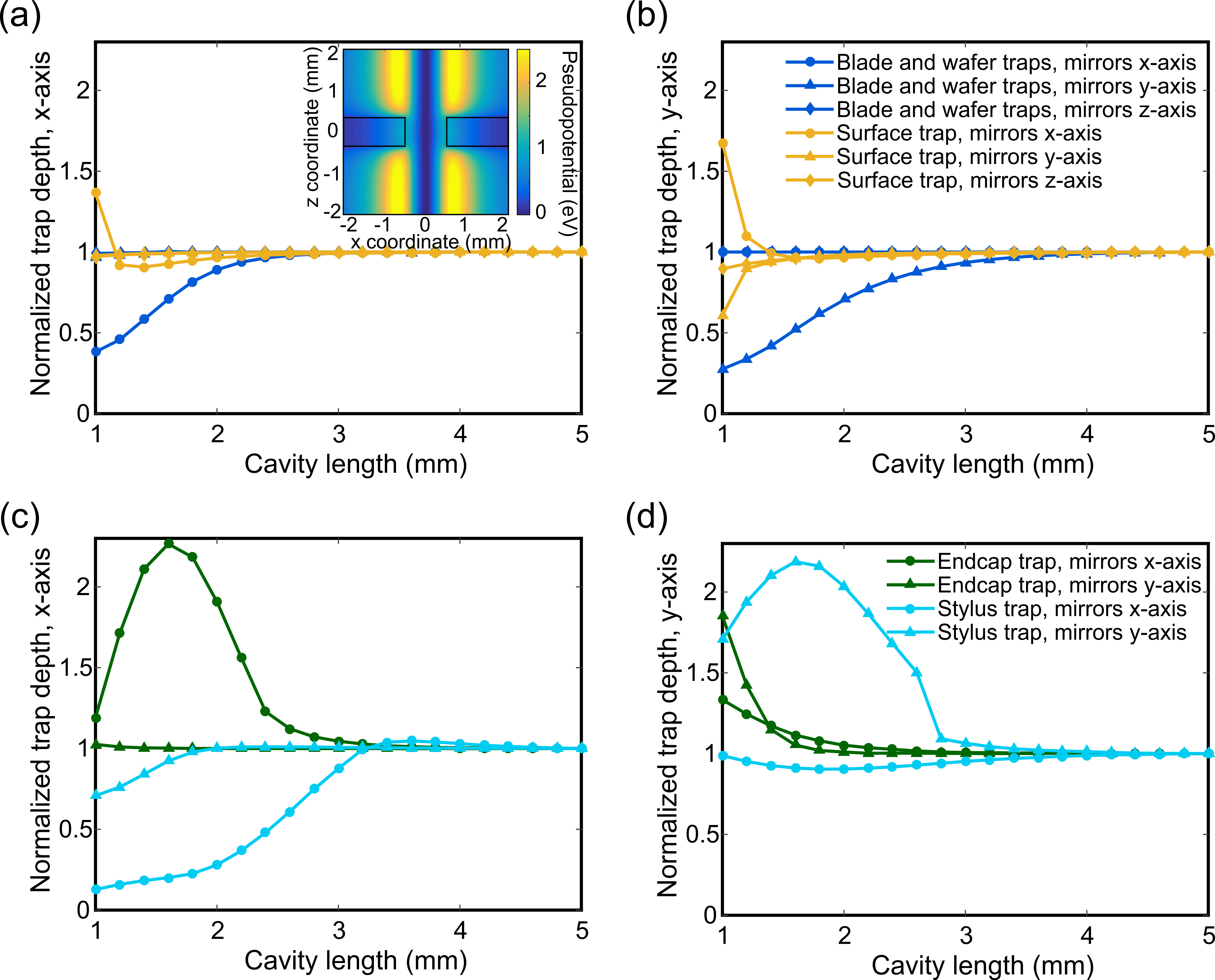}
  \caption{Trap depths calculated for different ion traps and different optical cavity orientations as a function of cavity length: (a) and (b) are the normalized trap depths in the $x$- and $y$-axis directions of the linear traps. The wafer and blade traps behave similarly and are thus represented by a single data set. The inset in (a) shows how the trapping field is deformed by the presence of the cavity in the blade trap at a cavity length of $1$  mm. It can be seen that the trap depth is significantly reduced in the $x$-direction due to the intersection of the cavity with the potential. (c) and (d) are the normalized trap depths in the $x$- and $y$-axis directions of the cylindrical traps. }
  \label{fig:barrier_data}
\end{figure*}

To characterize the effect of the cavity mirrors on the trapping potential we study the relative changes of the trap depths and secular frequencies.
 Figure~\ref{fig:barrier_data} shows changes of the trap depths as a function of the cavity length for different cavity alignments. We have normalized the trap depths with the values obtained in simulations without the optical cavities. The most notable result in Fig.~\ref{fig:barrier_data} is that the effect of the cavity mirrors is reduced if the cavity axis coincides with the symmetry axis of the trapping potential. The symmetry axes are along $z$- and $y$-axis for the linear and cylindrical traps respectively. For the blade and wafer linear traps, there is no noticeable degradation of the trap depth if the cavity axis is aligned along the $z$-axis.  However, when the cavities are aligned perpendicular to the $z$-axis, the trap depths drop below $50 \%$ of the unperturbed case at cavity lengths comparable to the inter-electrode distance ($\sim$ $1$ mm). Similarly, the trapping potential deformation is much smaller in the surface trap if the cavity is aligned along the trap axis compared to a perpendicular alignment.
Likewise in the endcap trap, when aligned along the $y$-axis, the effect of the cavity mirrors on the relative trap depths is smaller compared to when the cavity is aligned along the $x$-axis. Even though less pronounced, also in the stylus trap, the effect of the dielectric mirrors on the trapping field is significantly reduced if aligned along the symmetry axis ($y$-axis) of the trap electrodes.

The reduced distortion of the trapping field when the cavity axis coincides with the trap's symmetry axis can be easily understood by considering the behavior of the electric field on the mirror-vacuum interfaces. While the electric field component parallel to the surface of the dielectric is unaffected, the perpendicular component changes according to the boundary condition $E_{\mbox{\tiny dielectric}}^{\bot} = E_{\mbox{\tiny vacuum}}^{\bot}/\epsilon_{\mbox{\tiny dielectric}}$. This leads to a deflection of the electric field on the surface unless the electric field is either only perpendicular or parallel to the surface. In the case where the cavity axis is aligned along the symmetry axis of the trap, the electric field created by the electrodes is predominately perpendicular to the dielectric material of the mirrors. Thus, the deformation of the trapping field is minimum.

When the cavity alignment is perpendicular to the symmetry axis of the electrode configuration, the performance of the linear and endcap traps are very similar. Both traps exhibit a change in the trap depth starting from about $3$ mm cavity length (potential change $\approx 5\%$).
The stylus trap shows this deterioration at a similar distance but the drop in the trap depth is more rapid and becomes as low as $25\%$ at $2$ mm in contrast with the well above $50\%$ for the linear and endcap traps. This makes it the least robust ion trap when combined with dielectric cavities.
Due to the open geometry, the trapping field extends further out from the trap center compared to the other traps with closed structures. Therefore the overlap of the extending trapping field with the cavity mirrors is significantly larger even at long cavity lengths.

Surprisingly, the surface trap suffers far less from its open structure than the stylus trap. Even though both trap depths increase by adding the mirror above the electrode plane, the detrimental effect of the cavity when aligned perpendicular to the trap axis is significantly reduced in the surface trap.
Even compared with the blade/wafer and endcap traps, the effect of the cavity on the trapping potential seems significantly smaller at cavity lengths greater than $1.5$ mm . 
However, at short cavity lengths the disturbance increases significantly faster than for all the other traps. The reason for this is the close distance between the ion and the electrodes (300 $\mu m$) which provides effective shielding from the influence of the dielectric. 

\begin{figure*}[ht]
  \centering
  \includegraphics[width=13cm]{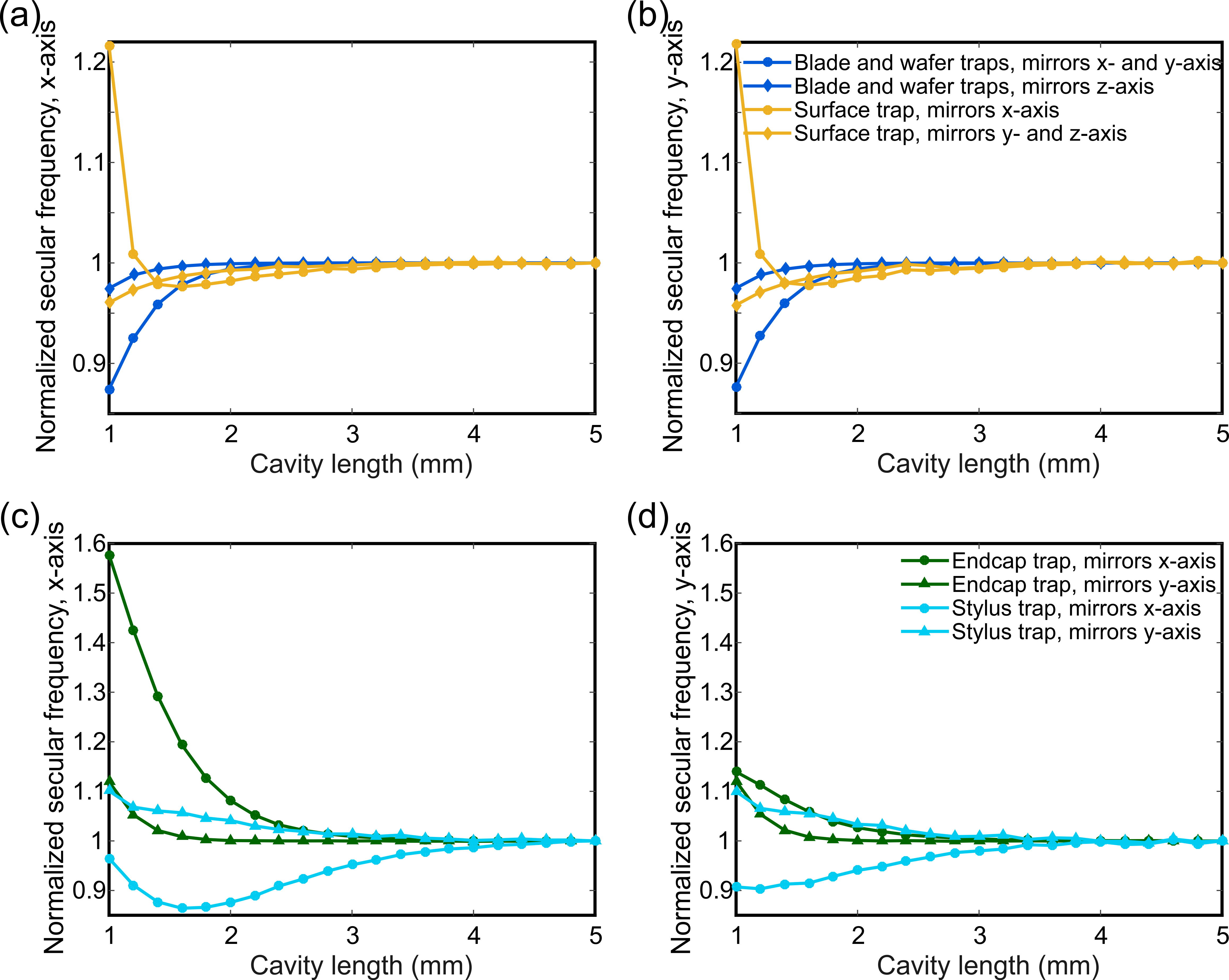}
  \caption{Secular frequencies calculated for different ion traps and different optical cavity orientations as a function of the caivty length: (a) and (b) are the normalized secular frequencies in the (a) $x$- and (b) $y$-axis directions of the linear traps. The wafer and blade traps behave very similarly and are thus represented by a single data set. (c) and (d) are the normalized secular frequencies in the (c) $x$- and (d) $y$-axis directions of the cylindrical traps.}
  \label{fig:freq_data}
\end{figure*}

To further examine the effect of the dielectric mirrors on the trapping field, we study relative changes of the secular frequencies for different traps and cavity alignments as a function of the cavity length (see Fig.~\ref{fig:freq_data}).
The frequencies are normalized by the values without the cavities as in Fig.~\ref{fig:barrier_data}.  
In all the configurations, except the endcap trap with mirrors along the $x$-axis, the secular frequencies change by less than 10\% at cavity lengths down to 1~mm, in contrast to the much larger relative changes of the trap depths discussed above.

\section{Cavity misalignment}
\label{sec:mirr}

\begin{figure*}[ht]
  \centering
 \includegraphics[width=14cm]{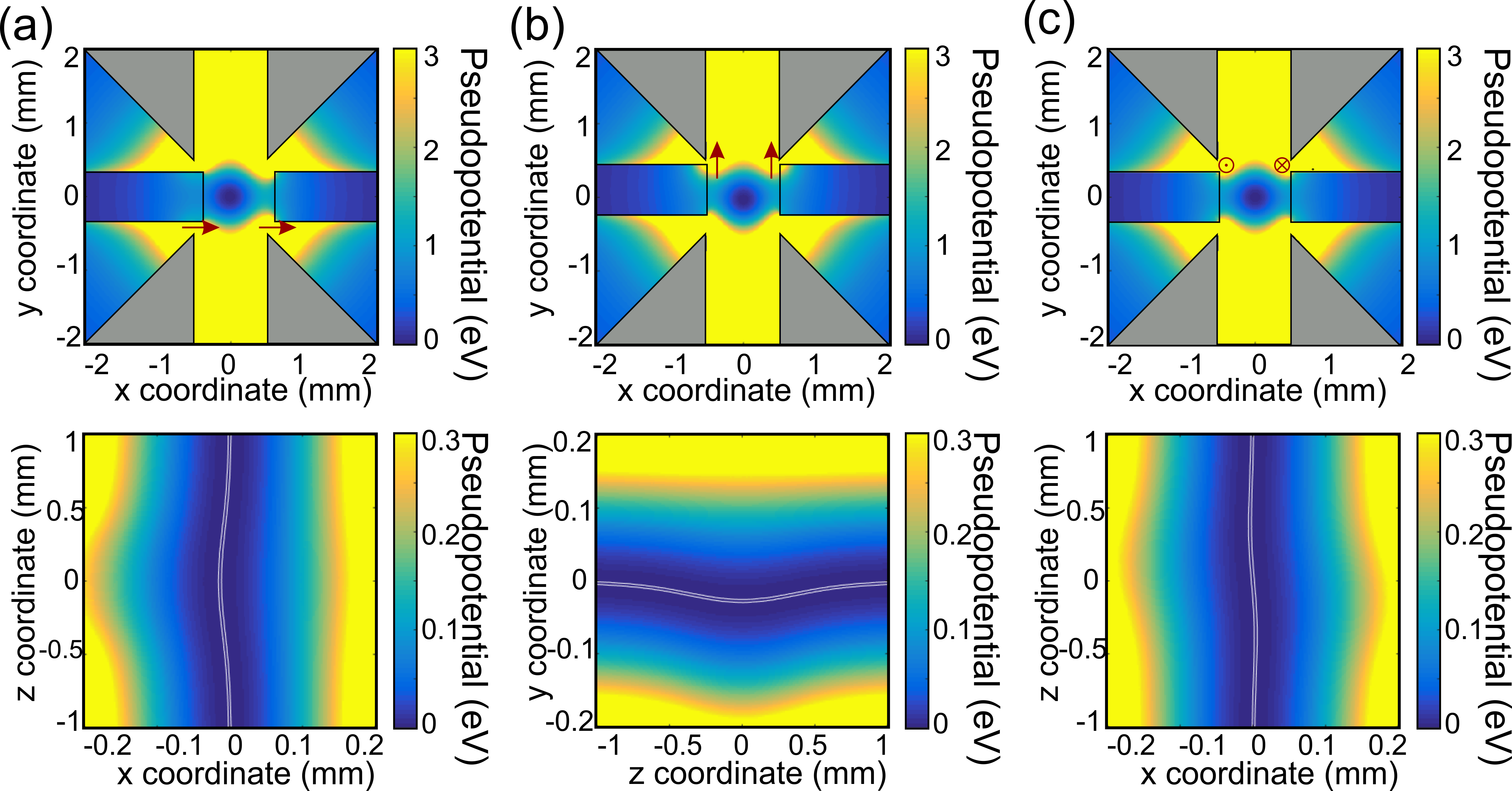}
  \caption{Configurations of cavity misalignment and the trapping potentials. }
  \label{fig:cavity_misalignment}
\end{figure*}

In Sec.~\ref{sec:dist-potent-cavity} we assumed that the cavity axis is perfectly aligned and the mirrors are positioned symmetrically around the trap center. In practice, however, this is not necessarily the case and often a small misalignment is observed. To investigate the effect of cavity misalignment on the trapping potential we focus on one specific ion trap, the linear blade trap. We have chosen this configuration as there has been an earlier experimental investigation of this effect with this trap geometry \cite{KellerThesis}.
We have tested three possible cavity orientations (shown in Fig.~\ref{fig:Mirr}). For each cavity orientation, the mirrors were shifted by $0.1$ mm in three different ways as illustrated in Fig.~\ref{fig:cavity_misalignment} for a cavity oriented along the $x$-axis: translational misalignment along (a) the longitudinal and (b) transverse direction, and (c) skewed misalignment in   the transverse direction. Perturbations to the trapping potential were calculated with a cavity length of $1$ mm.

When mirrors are aligned along the $x$- or $y$-axis and displaced along the cavity axis (Fig.~\ref{fig:cavity_misalignment}(a)), the trapping potential becomes asymmetric and the trap depth is reduced. The trap depth decreases by $20\%$ for the cavity oriented along the $x$-axis and by $30 \%$ for the cavity oriented along the $y$-axis. No reduction in the trap depth is observed with any type of misalignment when the cavity is aligned along the $z$-axis.

Any asymmetric mirror misalignment breaks the trap symmetry and hence shifts the location of a trapped ion. With a $1$ mm-long cavity oriented along the $x$-axis, the asymmetric mirror offset of $0.1$ mm in the $x$- and $y$-directions moves the ion position by $14$ $\mu$m and $27$ $\mu$m respectively from the unperturbed center in the direction opposite to the mirror shifts (see Fig.~\ref{fig:cavity_misalignment}(a) and (b)). When the cavity is oriented vertically, the ion position is shifted by $62$ $\mu$m and $14$ $\mu$m in the $x$- and $y$-directions for the mirror offsets in the corresponding directions. When the cavity is aligned along the $z$-axis, the ion displacements in the $x$- and $y$-directions are only $4$ $\mu$m and $6$ $\mu$m respectively. More insight on the trap distortion can be gained by looking at the way in which the potential is deformed. Figure \ref{fig:cavity_misalignment} shows the disturbed trapping potential for misaligned cavities. The offset of the cavity shifts the trap center towards the closest mirror. This displacement changes with the position along the trap axis. This effect is particularly pronounced if the mirrors are offset transversely  to the cavity axis and in opposite directions to each other (see Fig.~\ref{fig:cavity_misalignment}(c)). The resulting nodal line of the trapping field then exhibits a figure 'S' shape which can lead to an axial rf field component if the ion is displaced from the nodal line. Even though the ion's position can be adjusted in a way that it resides on the nodal line, the axial field components make it notoriously difficult to find external static fields to ensure that the ions is placed on the field's nodal line.
 
\begin{figure}[ht]
  \centering
  \includegraphics[width=6.5cm]{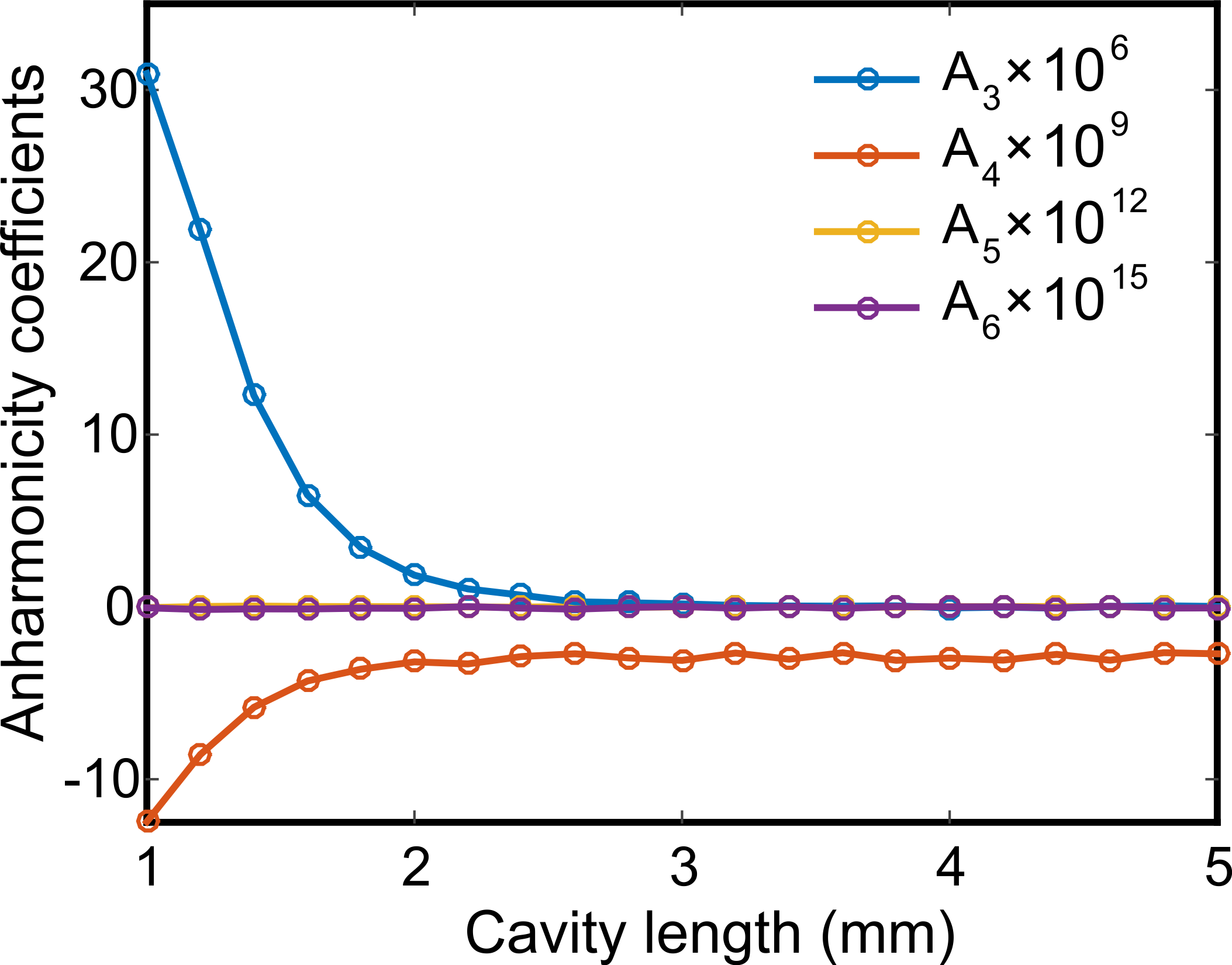}
  \caption{Anharmonicity of the trap potential due to a axial misalignment of the cavity mirrors.}
  \label{fig:anharmonicity}
\end{figure}

In addition, we have analyzed the anharmonicity of the trapping potential along the cavity axis for the case of Fig.~\ref{fig:cavity_misalignment}(a).
For this, the potential is expanded in a Taylor series 
\begin{equation}
\Phi_{\mbox{\tiny pseudo}}= \Phi_0 + \sum_{n=2}^6 C_n\cdot x^n,
\end{equation}
where $C_n$ are the expansion coefficients and $\Phi_0$ is an offset.
To ensure the expansion around the potential minimum, the linear expansion coefficient is forced to be zero. 
The suitable fitting range was determined by increasing the fitting range from $10$ $\mu m$ until the fitting error is below $10^{-6}$ and the fitting parameters are stabilized. This procedure results in a fitting range of 100 $\mu m$. To compare the fitting parameters of the Taylor expansion, we use the anharmonicity coefficients defined as a ratio between the anharmonic terms and the quadratic fitting parameter:
\begin{equation}
A_n = \frac{C_n \cdot l_0^{n-2}}{C_2}.
\nonumber
\end{equation}
The characteristic length scale $l_0$ is the width of the ions thermal wavepacket at the Doppler temperature in the trapping potential along the $x$-direction. Figure \ref{fig:anharmonicity} shows the anharmonicity coefficients for a range of cavity lengths between 1 mm and 5 mm. The characteristic length at the Doppler temperature of $590$ $\mu$K changes from $75$ nm to $64$ nm in this cavity length range due to the change in secular frequency.

\section{Effect of surface charges}
\label{sec:charges}

\begin{figure*}[ht]
  \centering
  \includegraphics[width=16cm]{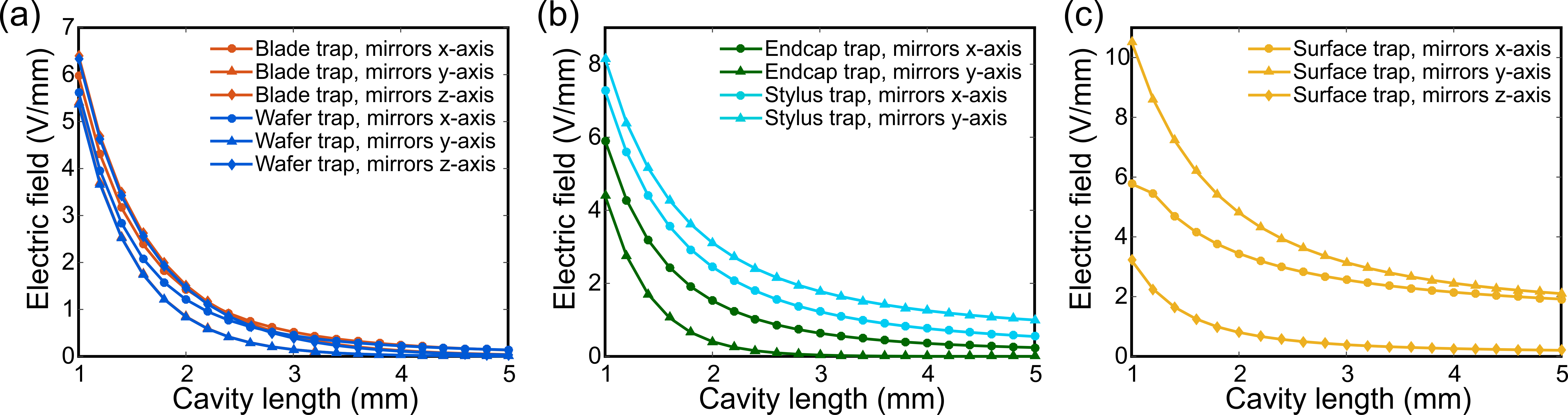}
	\caption{Electric field in the trap center generated by surface charges on one of the cavity mirrors in the case of (a) blade and wafer trap, (b) endcap and stylus trap, and (c) surface trap design.}
  \label{fig:SurfaceCharge}
\end{figure*}

We also investigate the effect of charges on the cavity mirrors on the trapping potential. In ion trap systems, the presence of laser radiation for generating and cooling ions are often in the UV region of the optical spectrum. Light scattering of these laser beams can cause charging either through direct interaction with the dielectric material \cite{Harlander} or through photo-electron emission which can accumulate on dielectric surfaces. While other dielectric materials in the ion trap can be put either far away from the trap center or shielded, this is not possible for the dielectric mirrors of short cavities. To simulate the effect of potential charging of the cavity mirrors,  we calculate the electric field at the trap center generated by a constant surface charge of $10^{-6} C/m^2$ on one of the cavity mirrors while the other mirror is kept uncharged. This is because charging both mirrors would, at least partially, cancel the effect on the trapped ions. 
We have simulated the static electric field at the trap center generated by the surface charge in the same configuration as in Sec.~\ref{sec:trap} (see Fig.~\ref{sec:trap}). The amplitudes of the simulated fields are shown in Fig.~\ref{fig:SurfaceCharge}.
While the electric field is almost independent of the cavity orientation for the blade trap, in the wafer trap the close vicinity of the rf electrodes to the cavity mirrors in the $y$-direction shields the charge and thus results in a reduced field at the trap center. In the case of $x$- and $z$-oriented cavity, the generated fields in the wafer trap are very similar to the blade trap.
The endcap trap with a cavity aligned along the $y$-axis shows the smallest electric field due to the shielding of the charge provided by the rf-electrodes. In contrast, the open structure of the stylus ion trap shows reduced shielding and a higher electric field in both cavity orientations. Very similarly, the surface trap (Fig.~\ref{fig:SurfaceCharge}(c)) provides the least shielding to show the highest static electric fields among all the tested geometries.

\section{Mirror material}
\label{sec:material}
 
\begin{figure*}[ht]
  \centering
  \includegraphics[width=14cm]{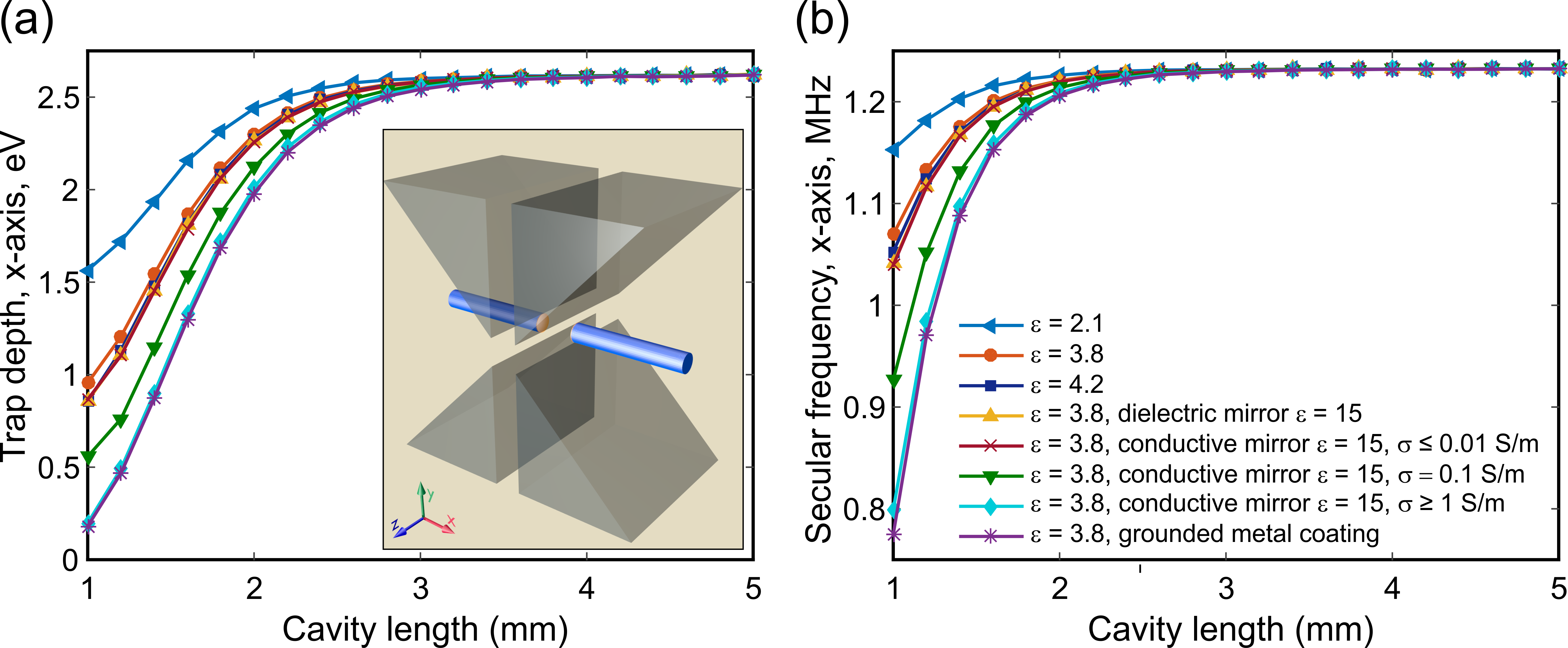}
  \caption{(a) Trap depth and (b) secular frequencies calculated for the blade linear trap with optical cavities with various mirror materials. The mirror materials include an additional layer of a high dielectric constant and conductive coatings on the mirror facets or on the entire substrates. Inset in (a) shows in the geometry of the ion trap. }
  \label{fig:Material}
\end{figure*}

Finally, we investigate the effect of the mirror material on trapping potential distortions using the blade linear trap. For this we have simulated trapping fields for mirrors with dielectric constants of $\epsilon = 2.1$ , 3.8 and 4.5 (typical range for silica glasses).  Figure~\ref{fig:Material} shows how the trap depth and secular frequency are affected by the presence of the mirrors. As expected, the trapping field deformation is larger for higher dielectric constants. For the dielectric constant of $\epsilon = 3.8$ we have also performed a simulation with an additional $10$ $\mu$m thick layer of a high dielectric constant ($\epsilon = 15$) material on the mirror facets in order to emulate the high reflective mirror coatings. Due to the small volume of this layer, the effect is insignificant (compare the red and yellow traces in Fig.~\ref{fig:Material}). 

In order to avoid a build up of electric charges on the dielectric surfaces, conductive layers may be deposited on the mirror surface.
We have investigated the influence of the conductivity of such a layer on the end facets of the mirror substrates by performing a full rf trapping field simulation, taking the time variation of the trapping field into account (see Fig.~\ref{fig:Material}).The trap depth deteriorates rapidly with increasing conductivity and saturates at high conductivities. The trap depth drops by more than 40\% when the conductivity increases from $0.01$ S/m to $1$ S/m but there is no significant change when the conductivity increases further. Figure \ref{fig:Material} also contains a full rf simulation of mirror substrates which are entirely covered by a grounded metal layer.
Even though this configuration is the best to prevent charging of the mirror substrates it causes the largest deterioration of the trapping potential.

\section{Conclusion}
\label{sec:conclusion}

We have numerically investigated the effect caused by the presence of a dielectric material on the trapping field in five different rf ion trap designs and cavity alignments. 
All traps were scaled to have typical electrode gaps of 1 mm for a fair comparison. It should be stressed that all the results can be easily rescaled to any trap dimensions depending on experimental needs. 
As a measure of the influence from the cavity mirrors, we have used the changes in the trap depth and the secular frequencies.   
Considering the deformation of the trapping field due to the dielectric mirrors (Sec.~\ref{sec:dist-potent-cavity}), the effect on the stylus traps are considerably larger in comparison with the other electrode configurations. In addition, the trapping field deformation is significantly reduced if the cavity axis is aligned with the symmetry axis of the electrode configuration e.g. the trap axis for linear traps.
Misalignment of the cavity mirrors with respect to the mirror substrate axis (transverse misalignment) or displacement of the cavity along its axis (longitudinal misalignment) results in a local shift of the trap center (Sec.~\ref{sec:mirr}). While these shifts still result in a field-free nodal line, the local deformation could lead to ion micro-motion along the trap axis. Even though this can be compensated, it may constitute an experimental complication. Asymmetric mirror positions also increase the anharmonicity of the trapping potential which can be associated with increased heating and parametric trap instabilities at higher ion temperatures.
The presence of charges on the dielectric surfaces causes static electric fields at the trap center which displace the ion from the rf-null point. Even though these static fields can be compensated by applying appropriate voltages to axillary or the rf electrodes, their fluctuations can impair localisation of the ions in the trap. Therefore, we have simulated the electric field at the trap center caused by surface charges on the mirrors (Sec.~\ref{sec:charges}). Similarly to the results in Sec.~\ref{sec:dist-potent-cavity}, the residual static electric fields in the trap center of a given surface charge is larger for open electrode structures.
To consider the effects of the mirror materials, we have conducted simulations for a specific trap geometry and cavity alignment (Sec.~\ref{sec:material}). For dielectric materials, the trapping field deformation increases with increasing dielectric constant, an expected behaviour. 
This effect significantly increases by increasing the mirror surface conductivity, for which we have used a full rf simulation of the trapping field. 

Despite several efforts, combining ion traps with optical cavities is still a significant challenge. Even though our simulations do not point towards a single specific effect which can explain the present difficulties in full, it clearly shows differences among different ion trap designs and mirror alignments in regards of the effect caused by the presence of the dielectric mirrors.
Closed electrode structures, such as the blade/wafer and endcap traps, with the cavity aligned along the symmetry axis of the electrode geometry exhibit the least deformation of the trapping field and the lowest effect from mirror charging. 

\section*{Acknowledgements}
We gratefully acknowledge support from EPSRC through the UK Quantum Technology Hub: NQIT - Networked Quantum Information Technologies (EP/M013243/1) and EP/J003670/1. The data reported in this work will be accessible at~\cite{data}.

\bibliography{References}

\end{document}